\documentclass[a4paper,11pt]{article}
\usepackage[dvipdfmx]{graphicx}

\usepackage{jheppub} 
\usepackage[T1]{fontenc} 

\title{\boldmath Aether SUSY breaking:\\
Can aether be alternative to F-term SUSY breaking?}

\author[a]{Yusuke Yamada}

\affiliation[a]{Research Center for the Early Universe (RESCEU), Graduate School of Science,\\ The University of Tokyo, Hongo 7-3-1,
Bunkyo-ku, Tokyo 113-0033, Japan}

\emailAdd{yamada@resceu.s.u-tokyo.ac.jp}

\preprint{RESCEU-13/21}

\abstract{We investigate supersymmetry (SUSY) breaking scenarios where both SUSY and Lorentz symmetry are broken spontaneously. For concreteness, we propose models in which scalar fluid or vector condensation breaks Lorentz symmetry and accordingly SUSY. Then, we examine whether such scenarios are viable for realistic model buildings.  We find, however, that the scalar fluid model suffers from several issues. Then, we extend it to a vector condensation model, which avoids the issues in the scalar fluid case. We show that accelerated expansion and soft SUSY breaking in matter sector can be achieved. In our simple setup, the soft SUSY breaking is constrained to be less than $\mathcal{O}(100)$TeV from the constraints on modification of gravity.}
\makeatletter
\gdef\@fpheader{}
\makeatother

\begin{document}
\maketitle
\flushbottom

\section{Introduction}
Supersymmetry (SUSY) has been a plausible candidate for the physics beyond the standard model of particle physics as well as general relativity. In particular, superstring theory can be a possible ultraviolet(UV) completion of SUSY including quantum gravity. SUSY plays an important role in string compactification, since the presence of unbroken SUSY ensures the stability of the vacuum. Nevertheless, in realistic model buildings, SUSY needs to be (spontaneously) broken at some scale. One of the reason for SUSY breaking is that collider experiments so far have not yet discovered any signature of expected SUSY partners of the standard model particles, which implies that superpartners have to obtain masses much larger than that of standard model particles. Besides that, from cosmological observation, our Universe is in an accelerated expansion phase, which also requires SUSY to be broken. Therefore, SUSY breaking mechanism is an important issue for realistic model buildings.

The most of models of spontaneous $\mathcal{N}=1$ SUSY breaking can be classified into either F-term or D-term breaking scenarios, where the auxiliary scalar components of SUSY multiplets get non-vanishing vacuum expectation values (VEV). In particular, F-term SUSY breaking scenario has been considered in various model buildings. For example, dynamical SUSY breaking~\cite{Witten:1982df,Affleck:1983mk,Affleck:1984xz,Affleck:1984uz,Izawa:1996pk,Intriligator:1996pu,Intriligator:2006dd} has a good theoretical control of the SUSY breaking scale, which would be necessary to realize the hierarchy between Planck scale $M_{\rm pl}$ and SUSY breaking scale. F-term breaking plays a crucial role in string theoretical realizations of de Sitter vacua such as KKLT~\cite{Kachru:2003aw} and LVS~\cite{Balasubramanian:2005zx,Conlon:2005ki}. In those models, anti-D3 brane breaks SUSY spontaneously, (not explicitly)~\cite{Kachru:2002gs}. The recent studies~\cite{McGuirk:2012sb,Kallosh:2014wsa,Bergshoeff:2015jxa,Aparicio:2015psl,Vercnocke:2016fbt,Kallosh:2016aep,GarciadelMoral:2017vnz,Aalsma:2018pll,Cribiori:2019hod} show that anti-D3 brane can be effectively described by a non-linear realization of SUSY~\cite{Volkov:1972jx,Volkov:1973ix,Rocek:1978nb,Ivanov:1978mx,Lindstrom:1979kq,Casalbuoni:1988xh,Komargodski:2009rz}, which necessarily has a non-vanishing F-term. The other scenario is D-term breaking models, where the auxiliary field in gauge multiplets get non-vanishing vacuum expectation values.

What else is possible as spontaneous SUSY breaking scenarios? Although the F-term or D-term breaking scenarios provide building blocks for realistic models, it would be interesting to think about other options if possible. In this work, we propose an alternative scenario, which we dub as {\it aether SUSY breaking} where SUSY is spontaneously broken by vector condensation rather than non-vanishing F-terms or D-terms. Such a scenario leads to Lorentz violation, which would show phenomenological features that differ from F-term and D-term breaking models. As a prototype of the scenario, we first discuss kinetic SUSY breaking scenario where scalar field fluid breaks SUSY, but one immediately finds several problems, which seem to make difficult both ultraviolet completion and phenomenologically consistent model buildings. Then, we discuss how to cure the issues, which leads us to a vector condensation scenario rather than one with scalar fluid.\footnote{The appearance of the preferred frame vector field is also known in the context of Einstein-Aether model~\cite{Jacobson:2000xp,Jacobson:2008aj}.}

Let us clarify the idea behind our models: There are some symmetry breaking patterns within SUSY. Without gravity, F-term / D-term breaking is the case that SUSY is spontaneously broken while the Lorentz symmetry is left unbroken. The positive vacuum energy does not have any meaning unless (super)gravity is turned on, and therefore, within global SUSY, these breaking scenarios realize SUSY broken Minkowski spacetime. Once gravity is turned on, and if total vacuum energy density is positive, the Universe experiences the de Sitter phase and de Sitter isometries become the remaining spacetime symmetry. If by chance the total energy is absolutely zero, we find Minkowski spacetime with broken SUSY within supergravity. On the other hand, if time or spatial translation is broken, some part of Lorentz symmetry and SUSY would also be broken simultaneously. In kinetic SUSY breaking model, the time translation, Lorentz symmetry and SUSY are broken simultaneously. The difference is that in kinetic SUSY breaking, these symmetries are broken even without gravity, whereas the F-term breaking only breaks SUSY. In aether SUSY breaking model, vector condensation breaks Lorentz symmetry.

Recently the swampland conjectures, which are claimed to be the conditions on low energy field theories consistent with quantum gravity, are actively discussed (for a comprehensive review of swampland conjectures, see e.g.~\cite{Palti:2019pca}). In particular, one of the conjectures says that quantum gravity cannot realize de Sitter vacua~\cite{Obied:2018sgi,Garg:2018reu,Ooguri:2018wrx}. Although those conjectures are controversial, it may suggest that new possibilities of accelerated expansion are necessary. Since SUSY breaking and realization of accelerated expansion are tightly related to each other within SUSY models, alternative realizations of accelerated expansion may also imply necessity of new types of SUSY breaking scenarios. In this sense, our proposal may open new possibilities for realistic model buildings from UV theories such as string theory.

Independently of the UV completions, our new models of SUSY breaking have an interesting phenomenological feature: The spontaneous breaking of the time translation invariance modifies the property of gravitational degrees of freedom. As a result, we find that SUSY breaking scale, which is related to time translation violation scale, is constrained by tests of general relativity rather than collider experiments. Regarding collider experiments, at least one should realize the splitting of the mass scale between that of the standard model sector and its superpartner sector. We discuss how vector condensation can realize soft SUSY breaking. Interestingly, due to the constraints on the modification of gravity, the soft SUSY breaking scale within our simple model is constrained to be less than ${\cal O}(100){\rm TeV}$, which can be either confirmed or excluded by future collider experiments.

The rest of the paper is organized as follows. In Sec.~\ref{rev}, we briefly review F-term SUSY breaking scenario for comparison with our new scenarios. We propose the kinetic SUSY breaking scenario in Sec.~\ref{kineticsb}, which explains the main idea of our proposal. However, we immediately find several issues in the model as discussed in Sec.~\ref{problem}. Then, we propose the aether SUSY breaking scenario as a more realistic model in Sec.~\ref{Ether}. We also discuss constraints on the characteristic vector condensation scale, which is intimately related to soft SUSY breaking scale. Finally, we conclude in Sec.~\ref{concl}.

\section{Brief review of F-term SUSY breaking}\label{rev}
Let us briefly review F-term SUSY breaking scenarios. First, we discuss a general aspect of SUSY breaking. Suppose a chiral superfield is responsible for SUSY breaking. For simplicity, we will use superfields in global SUSY. A chiral superfield is given by~\footnote{We will use the notation of~\cite{Wess:1992cp} throughout this paper.}
\begin{equation}
    \Phi(y,\theta)=A(y)+\sqrt{2}\theta\psi(y)+\theta\theta F(y),
\end{equation}
where $A(x)$, $\psi_\alpha(x)$, and $F(x)$ are a complex scalar, a Weyl fermion, and a complex auxiliary scalar field (F-term), respectively. Here, we have used the chiral coordinate $y^\mu=x^\mu+{\rm i}\theta^\alpha(\sigma^\mu)_{\alpha\dot\alpha}\bar\theta^{\dot\alpha}$, where $\alpha,\dot\alpha$ denote two-component spinor indices.
We can read off the condition on SUSY breaking from the transformation law of $\psi$, which is given by
\begin{equation}
    \delta\psi_\alpha ={\rm i}(\sigma^\mu)_{\alpha\dot\alpha}\bar\epsilon^{\dot\alpha}\partial_\mu A+\sqrt{2}\epsilon_\alpha F,
\end{equation}
where $\epsilon_\alpha$ and $\bar{\epsilon}^{\dot\alpha}$ denote SUSY transformation parameters. One can see that $\langle \delta\psi\rangle\neq 0$ when either $\langle F\rangle\neq0$ or $\langle\partial A\rangle\neq0$. The former situation is called the F-term SUSY breaking. Most of the phenomenological models are based on such a scenario, whereas, to our best knowledge, the latter situation has not been applied to realistic model buildings.~\footnote{Non-vanishing D-term in gauge sector can also break SUSY spontaneously. Recently, D-term breaking models with a new class of Fayet–Iliopoulos term has been proposed~\cite{Cribiori:2017laj,Antoniadis:2018cpq,Kuzenko:2018jlz}. Here, however, we will focus only on F-term breaking.} In later sections, we propose a model based on the latter situation.

A simple F-term breaking scenario can be briefly summarized as follows: In supergravity, the (F-term) scalar potential is given by
\begin{equation}
    V_F=e^{K/M_{\rm pl}^2}\left(D_IWK^{I\bar{J}}D_J\bar{W}-3\frac{|W|^2}{M_{\rm pl}^2}\right),\label{F1}
\end{equation}
where $M_{\rm pl}\sim2.4\times 10^{18}{\rm GeV}$ denotes the Planck scale, $K(\Phi^I,\bar{\Phi}^{\bar{J}})$, $W(\Phi^I)$ denote K\"ahler and superpotential, respectively, and $\Phi^I$ ($\bar{\Phi}^{\bar{J}}$) are (anti-)chiral superfields. The K\"ahler covariant derivative is defined as $D_{I}W=\partial_I W+K_IW/M_{\rm pl}^2$, $K^{I\bar{J}}$ is the inverse of the K\"ahler metric $K_{I\bar{J}}=\partial_I\partial_{\bar{J}}K$ and $\partial_I$ denotes the derivative with respect to scalars $\Phi^I$. The quantities characterizing the SUSY breaking are the VEV of F-terms $\langle F^I\rangle$ and the gravitino mass $m_{3/2}$, which are given by
\begin{eqnarray}
    F^I&=&-e^{K/(2M_{\rm pl}^2)}K^{I\bar{J}}D_{\bar{J}}\bar{W},\\
    |m_{3/2}|^2&=&e^{K/M_{\rm pl}^2}\frac{|W|^2}{M_{\rm pl}^4}.
\end{eqnarray}
In terms of $F^I$ and $m_{3/2}$, we may rewrite the scalar potential as
\begin{equation}
    V_F=F^IK_{I\bar{J}}\bar{F}^{\bar{J}}-3|m_{3/2}|^2M_{\rm pl}^2.\label{F2}
\end{equation}
In order to achieve (1) SUSY breaking, (2) accelerated expansion of the Universe, and (3) non-vanishing gravitino mass, we need the situation where $\langle F^IK_{I\bar{J}}\bar{F}^{\bar{J}}\rangle-3|m_{3/2}|^2M_{\rm pl}^2\sim \Lambda^4\sim \mathcal{O}(10^{-120}M_{\rm pl}^4)$, and $|m_{3/2}|^2\neq0$ where $\Lambda$ denotes the cosmological constant. This requirement implies $\langle F^I\rangle\neq0$ for some chiral superfield $\Phi^I$ and $\langle W\rangle\neq0$. If a single chiral superfield $S$ is responsible for SUSY breaking, we find the relation
\begin{equation}
    |F^S|\sim \sqrt{3}|m_{3/2}|M_{\rm pl}.\label{fg}
\end{equation}

The SUSY breaking effect is somehow mediated to matter sector, which realizes the mass splitting between the standard model particles and their SUSY partners. For instance, the higher-order K\"ahler mixing $\delta K=-\frac{1}{M^2}|S|^2|Q|^2$ reads to the soft mass term for a scalar field $Q$ as $m_Q^2=|\langle F^S\rangle|^2/M^2\sim 3|m_{3/2}|^2(M_{\rm pl}^2/M^2)$, where $S$ is the SUSY breaking multiplet with $\langle F^S\rangle\neq0$ and $M$ denotes some mass scale suppressing the higher-order term. This is often referred as gravity mediation especially for $M\sim M_{\rm pl}$.\footnote{Even if there is no higher-order K\"ahler mixing term, the models with minimal K\"ahler potential $K=|Q|^2+|S|^2$ can give a soft mass to the scalar $Q$ within supergravity. In general, however, the quartic K\"ahler term equally contributes to soft scalar mass.} There are other mediation mechanisms, but as we will show, our new scenario is similar to  gravity mediation type scenarios (for a review of other mediation models, see, for example,~\cite{Martin:1997ns}). For gauge sector, a gaugino may obtain a soft mass term $m_g\sim \langle F^S\rangle\partial_S\langle f \rangle/\langle f\rangle$ if we assume $S$-dependent gauge kinetic function $f(S)$. Even if we do not assume such a $S$-dependence on gauge kinetic functions, anomaly mediation induces gaugino masses~~\cite{Randall:1998uk,Giudice:1998xp,Bagger:1999rd}.

We stress that the F-term SUSY breaking is caused by non-vanishing scalar quantities, and therefore, the rest of spacetime symmetry, diffeomorphism is unbroken. The models we propose are very different in this respect, since we consider spontaneous breaking of spacetime symmetry other than SUSY.

\section{kinetic SUSY breaking}\label{kineticsb}
We consider an alternative scenario, which we dub as {\it kinetic SUSY breaking}. Unfortunately, this scenario itself faces several issues as we will discuss; however, those issues can be resolved in an extended model, {\it aether SUSY breaking}. Nevertheless, the main idea of aether SUSY breaking scenario is the same as that of kinetic SUSY breaking. Therefore, we will discuss kinetic SUSY breaking scenario somewhat in detail. 

As we discussed, SUSY breaking takes place either by VEV of F-terms or background with non-vanishing derivatives of scalar components. In the following, we consider the latter possibility. We should note that such a possibility itself has been pointed out in the literature~\cite{Koehn:2012te,Nitta:2017yuf,Gudnason:2018nof}; however, to our best knowledge, soft SUSY breaking mediation to matter sector as well as accelerated expansion of the Universe have not been addressed within such a framework.\footnote{The SUSY breaking scenario similar to ours was discussed in~\cite{Katz:2006rx}, where the authors imposed shift symmetry of scalar fields by applying differential operators, and therefore, they concluded that the soft SUSY breaking mediated to visible sector is very small. As we will show this is not the case in our setup.}

From a phenomenological viewpoint, we require a model to achieve the mass splitting between SUSY particles and the standard model particles, and accelerated expansion of the Universe, at least. Of course, these are not sufficient conditions for realistic SUSY models, but the above requirements are necessary to make a model consistent with current experiments and observations. The simplest realization of accelerated expansion is a positive cosmological constant, which is realized by the balance between F-term and gravitino mass term in F-term SUSY breaking scenario. However, if we consider quintessential models, positive cosmological ``constant'' is not necessary. 

For an illustrative purpose, we consider a toy model where the accelerated expansion is achieved by derivative terms of a scalar field, instead of (F-term) scalar
potential, which is known as k-essence~\cite{Chiba:1999ka}/ghost condensation~\cite{ArkaniHamed:2003uy}. To realize such a scenario, we consider the following SUSY Lagrangian~\cite{Khoury:2010gb},
\begin{equation}
    \mathcal{L}=\int d^4\theta\left( K+\frac{1}{16}T(\hat{X})D^\alpha\Phi D_\alpha\Phi\bar{D}_{\dot\alpha}\bar{\Phi}\bar{D}^{\dot\alpha}\bar{\Phi}\right),\label{PX}
\end{equation}
where $\Phi$ is the SUSY breaking chiral superfield, $K$ is a K\"ahler potential and $\hat{X}\equiv\frac{1}{4} (\partial(\Phi-\bar{\Phi}))^2$, and $D_\alpha$ and $\bar{D}_{\dot\alpha}$ denote spinor derivatives. We call the scalar component of $\Phi$ as $\Phi|_{\theta=0}=\frac{1}{\sqrt{2}}(\chi+{\rm i}\phi)$. We choose the K\"ahler potential to be $K=K(\Phi+\bar{\Phi})$ so that the Lagrangian has shift symmetry $\phi\to\phi+c$ where $c$ is a constant, and we identify $\phi$ as the k-essence scalar field. We note that, although this expression is for a global SUSY case, the (conformal) supergravity extension is straightforward, and there is no crucial difference for components (see e.g.~\cite{Koehn:2012ar,Koehn:2012te,Aoki:2014pna,Aoki:2015eba}). In the absence of superpotential, the equation of motion of the F-term $F^\Phi$ has the solution $F^\Phi=0$. Then, the on-shell action is given by 
\begin{equation}
    \mathcal{L}=-\frac12 K_{\Phi\bar{\Phi}}\left((\partial\phi)^2+(\partial\chi)^2\right)+T(\tilde{X})\left[\tilde{X}^2+\frac14 (\partial\chi)^4+\tilde{X}(\partial\chi)^2+(\partial\phi\partial\chi)^2\right],
\end{equation}
where $\tilde{X}=-\frac12(\partial\phi)^2$. Let us assume for now that $\chi=0$.\footnote{In this Lagrangian, this is a consistent solution of the equation of motion. Later, we will show that it is possible to add effective mass to $\chi$, which can also stabilize $\chi$.} We take the K\"ahler potential to be $K=\frac12(\Phi+\bar{\Phi})^2$. Let us consider a homogeneous background $\phi=\phi(t)$, and then the effective Lagrangian becomes
\begin{equation}
    \mathcal{L}=X+T(X)X^2\equiv P(X),
\end{equation}
where $X=\frac12\dot{\phi}^2$ and dot denotes the time derivative. By choosing an appropriate function $T(X)$, this model can realize effective de Sitter phase as we will discuss below.

As for other sector, we consider the standard setup,
\begin{equation}
    \mathcal{L}=\int d^4\theta K_0(\Phi^I,\bar{\Phi}^{\bar{J}})+\left(\int d^2\theta W(\Phi^I)+{\rm h.c.}\right),
\end{equation}
where $\Phi^I$ denotes both matter sector and moduli fields if exists. Here we emphasize that we will not consider F-term breaking scenarios. With vanishing F-terms, the F-term scalar potential~\eqref{F2} becomes
\begin{equation}
    V_F=-3|m_{3/2}|^2M_{\rm pl}^2.
\end{equation}
If R-symmetry is unbroken, this term should vanish as well, but then gravitino becomes massless. Here, we consider explicit breaking of R-symmetry by simply adding a constant to the superpotential, $\langle W\rangle=W_0={\rm const}$. In string moduli stabilization models, such as KKLT~\cite{Kachru:2003aw} and LVS~\cite{Balasubramanian:2005zx}, there appears non-vanishing (effectively) constant superpotential due to the remnant of flux stabilization of moduli fields and so on.  Without the k-essence sector $\Phi$, the scalar potential gives a negative constant,
\begin{equation}
    \langle V_F\rangle=-\frac{3|W_0|^2}{M_{\rm pl}^2},
\end{equation}
which means that the minimum realizes SUSY anti-de Sitter spacetime. Of course, this situation cannot describe our Universe, since the spacetime curvature is negative and also SUSY is unbroken.

Let us discuss how the accelerated expansion as well as SUSY breaking and its mediation to matter sector can be achieved. Assuming all but the k-essence scalar are stabilized, the effective action becomes
\begin{equation}
    S=\int d^4x\sqrt{-g}(P(X)+3|m_{3/2}M_{\rm pl}|^2),
\end{equation}
where the latter term in parentheses is a positive constant (= a negative cosmological constant). We take the background metric to be flat Friedmann-Robertson-Walker metric, $ds^2=-dt^2+a^2(t)d{\bf x}^2$, where $a(t)$ is the scale factor.
The equations of motion read
\begin{eqnarray}
&&(P_X+2P_{XX}X)\dot{X}+6HP_XX=0,\label{scalar}\\
&&3M_{\rm pl}^2H^2=\rho,\label{hubble}\\
&&\rho=2XP_X-(P+3|m_{3/2}M_{\rm pl}|^2),\label{rho}\\
&&p=P+3|m_{3/2}M_{\rm pl}|^2\label{p},
\end{eqnarray}
where $P_X=\partial_XP$, $P_{XX}=\partial_X\partial_XP$, $H=\dot{a}/a$, and $\rho$ and $p$ are energy density and pressure, respectively. We find that when $P_X=0$, namely, if there is a local minimum for $P(X)$ at $X=X_0={\rm const}$, the equation of motion~\eqref{scalar} is satisfied. Furthermore, the equation of state parameter becomes $w=p/\rho=-1$. If $\rho=-(P(X_0)+3|m_{3/2}M_{\rm pl}|^2)>0$ the accelerated expansion of the Universe takes place. From current observational data, the Hubble parameter is negligibly small but positive $H^2_0\sim \mathcal{O}(10^{-120}M_{\rm pl}^2)$, and this requires a miraculous cancellation between the k-essence term $-P(X_0)$ and the negative cosmological constant originating from other sector $-3|m_{3/2}M_{\rm pl}|^2$. This requirement seems unnatural, but the same is required for F-term uplifting cases, where the cancellation between positive and negative contribution in the F-term potential is realized by fine-tuning.

Let us discuss the condition on the k-essence term at the local minimum $P(X_0)$. Neglecting the tiny Hubble parameter, \eqref{hubble} implies
\begin{equation}
   3|m_{3/2}M_{\rm pl}|^2\sim -P(X_0)>0.
\end{equation}
Therefore, at the local minimum, $P(X_0)$ needs to be negative and slightly larger than the gravitino mass term $|P(X_0)|>3|m_{3/2}M_{\rm pl}|^2$, although the difference would be negligible. If these requirements are satisfied, the de Sitter phase of the Universe can be realized.

Thus far, we have discussed the accelerated expansion, but how about SUSY breaking mediation to matter sector? Even though F-term of any sector is vanishing, soft terms of visible sector can be realized as follows: For chiral matter sector, we may consider K\"ahler potential term
\begin{equation}
    \delta K=-\frac{1}{\tilde{M}^2}(\Phi+\bar{\Phi})^2|Q|^2,\label{softc}
\end{equation}
where $Q$ denotes a generic matter scalar field such as squarks and sleptons, and $\tilde{M}$ denotes an unknown mass parameter that suppresses the higher-order K\"ahler term. Such a higher-order term gives rise to the effective ``soft mass'' term
\begin{equation}
   \delta\mathcal{L}=-\frac{X_0}{\tilde{M}^2}|Q|^2,\label{soft}
\end{equation}
as the usual gravity mediation. This possibility was pointed out in~\cite{Farakos:2021mwt}, in which the author considered the slow-roll quintessence model, and such a mass term is negligibly small. On the other hand, we are now considering a non-slow-roll type quintessence model, and therefore, this contribution is no longer small. We also note that the similar quartic K\"aher potential term can be introduced to $\Phi$ itself as 
\begin{equation}
    \delta K=-\frac{1}{48\Lambda^2}(\Phi+\bar{\Phi})^4,
\end{equation}
where $\Lambda$ is a mass parameter. Such a term leads to the effective mass to $\chi$ 
\begin{equation}
    \delta\mathcal{L}\sim -\frac{X_0}{2\Lambda^2}\chi^2,
\end{equation}
which stabilizes $\chi$ at the origin and therefore our assumption $\chi=0$ can be consistently realized. We note that the shift symmetry of ${\rm Im}\Phi$ direction is important to suppress the mass term for the quintessence field $\phi$.

Let us turn to discussion on the soft SUSY breaking scale. In order to make an estimate on the value of the soft term, we parametrize $P(X)$ as
\begin{equation}
    P(X)=M^4R(Y)
\end{equation}
where $M$ denotes a mass parameter characterizing the higher dimensional operators in $P(X)$ and $Y=X/M^4$. The value of $Y_0=X_0/M^4$ depends on the choice of the function $P(X)$. But let us assume $Y_0\leq 1$. $M$ may be understood as a cut-off scale of this effective theory, so the value of $Y_0$ should be smaller than one, such that higher order terms becomes less important. This situation would be naturally realized in the ghost condensation scenario.\footnote{If we consider ghost condensate scenario, there appear two ghost scalars due to SUSY. The second ghost can be stable by some modification. See~\cite{Khoury:2010gb} for more details. } For k-essence scenario, some numbers of terms including higher-order terms need to be relevant to make a local minimum, which may seem unnatural from effective field theory viewpoint. For a concrete example, $R(Y)=Y+0.5Y^2-9Y^3+9Y^4$ has a local minimum at $Y_0\sim 0.64$ and $R(Y_0)\sim-0.005$. Of course, we show this example for an illustrating purpose, and the rest of discussions does not much depend on the choice of $P(X)$. One may find more reasonable models from UV theory. In either way, we will assume that there is a local minimum, where $X\neq0$ in the following. Suppose $|R(Y_0)|\sim \mathcal{O}(1)$, for simplicity, and then, since $|P(X_0)|\sim 3|m_{3/2}M_{\rm pl}|^2$ is required, we find the relation
\begin{equation}
    M\sim |m_{3/2}M_{\rm pl}|^{1/2},\label{Mrel}
\end{equation}
and this means that the UV scale $M$ is much larger than SUSY breaking scale $\sim|m_{3/2}|$. For simplicity, we assume $Y_0\sim 1$\footnote{$Y_0\ll1$ is more appropriate situation for a controllable effective field theory. However, we choose $Y_0\sim1$ for estimation.}, and then the value of the velocity $X_0$ is 
\begin{equation}
    X_0\sim M^4\sim |m_{3/2}M_{\rm pl}|^2.
\end{equation}
Thus, in this situation, we find the relation between $X_0$ and gravitino mass scale $|m_{3/2}|$. This is similar to the F-term breaking scenario where $|F^S|$ is related to the gravitino mass scale via the similar relation, see \eqref{fg}. In this case, the ``soft mass''~\eqref{soft} for matter fields is 
\begin{equation}
    m_{Q}^2\sim \frac{M_{\rm pl}^2}{\tilde{M}^2}|m_{3/2}|^2.
\end{equation}
When $\tilde{M}\sim M_{\rm pl}$, $m_{Q}^2\sim |m_{3/2}|^2$, and this is indeed similar to the case of the F-term SUSY breaking scenario.

As for gauge sector, anomaly mediation~\cite{Randall:1998uk,Giudice:1998xp,Bagger:1999rd} would give rise to the gaugino mass, since conformal compensator superfield, which is auxiliary degree of freedom, has non-vanishing F-term even though any physical chiral superfields have vanishing F-term~\cite{DEramo:2012vvz,DEramo:2013dzi}. The F-term of compensator $S_0$ is $|F^{S_0}|\sim |m_{3/2}|$ as usual and the gaugino is $m_{g}\sim \frac{g^2b_0}{16\pi^2}m_{3/2}$, where $g$ denotes a gauge coupling and $b_0$ the 1-loop beta function coefficient. Thus we have found that the kinetic SUSY breaking realizes the soft SUSY breaking in matter sector in a very similar manner to the F-term SUSY breaking scenario, by which realistic mass splitting between standard model particles and SUSY particles can be realized without conflicting the collider experiments so far.

\subsection{Problems of kinetic SUSY breaking}\label{problem}
We have seen that the kinetic SUSY breaking scenario may work as an alternative to F-term SUSY breaking scenario. However, there is an issue unique to the kinetic SUSY breaking scenario.

Let us show a problematic relation between the field excursion range of $\phi$ and the soft SUSY breaking mass scale. In kinetic SUSY breaking, the current inflationary stage is driven by a scalar field velocity $X=\frac12 \dot{\phi}^2$. In our simple solution $X=X_0={\rm const}$, we can estimate the field excursion to realize $\mathcal{O}(1)$ e-folding number as follows: The number of e-folding $N$ is related to the field excursion range as
\begin{equation}
    N=\int Hdt=\int d\phi\frac{H}{\dot{\phi}}\sim \frac{H_0}{\sqrt{X_0}}\Delta\phi,
\end{equation}
where $\Delta\phi$ is the field excursion range required for $N$ e-folding. On the other hand, from \eqref{soft}, $X_0$ is related to a soft mass $m_{Q}^2$, and therefore, we find the following relation
\begin{equation}
    \Delta\phi\sim \frac{m_Q}{H_0}N\tilde{M}.\label{exc}
\end{equation}
Since $H_0\sim \mathcal{O}(10^{-60}M_{\rm pl})$ is extremely small whereas $m_Q>\mathcal{O}(1){\rm TeV}\sim 10^{-15}M_{\rm pl}$, this relation means that the field excursion required in our model would be in general much larger than Planck scale $M_{\rm pl}$, unless we consider very unnatural parameter choices. For example, if $\tilde{M}\sim M_{\rm pl}$, $\Delta\phi>10^{45}M_{\rm pl}$, which is in strong tension with the distance conjecture~\cite{Ooguri:2006in}. Therefore, kinetic SUSY breaking cannot realize a model consistent with such a conjecture. The relation~\eqref{exc} is rather general in our scenario independently of details of k-essence scenarios. We also note that the relation~\eqref{exc} is also independent of the condition $-P(X_0)-3|m_{3/2}M_{\rm pl}|^2\simeq 3H_0^2M_{\rm pl}^2$, and therefore, modification of this relation cannot help to reduce the required excursion range. This result can be easily understood by a simple reason: The mass term is provided by scalar velocity $|\dot\phi|$, which needs to be sufficiently large to make mass splitting between SUSY particles and standard model ones. This splitting needs to last for about the age of our Universe, which requires very large excursion for $\phi$. Therefore, the kinetic SUSY breaking scenario inevitably faces this problem. Note that, if the shift symmetry $\phi$ enjoys is exact, the large field excursion might be not a problem. However, there are several discussions about the absence of global symmetry in quantum gravity~\cite{Kallosh:1995hi,Banks:2010zn} or the absence of global symmetry is proven in the context of AdS/CFT~\cite{Harlow:2018jwu,Harlow:2018tng}. Therefore, such a large field excursion might be impossible to realize within UV theories.

Other problem is model-dependent and associated with $P(X)$ models with $P_X(X_0)=0$. This situation leads to vanishing speed of sound for the fluctuation of $\phi$. In ghost condensation, it is pointed out that higher-order derivative terms provide $M^{-2}(\partial_i^2\phi)^2$ as leading spatial derivatives, which give the dispersion relation $\omega^2=k^4/M^2$~\cite{ArkaniHamed:2003uy}. Coupling to gravity leads to negative contribution to the right-hand side of the dispersion relation, and one finds Jeans-type instability, which leads to the constraint $M<100{\rm GeV}$, so that the cosmic microwave background is not smeared out~\cite{ArkaniHamed:2005gu}. If we also require the same constraint, we cannot realize $m_Q>1{\rm TeV}$ unless we assume extremely small $\tilde{M}$. Also, ghost condensation scenario may suffer from caustic formations~\cite{ArkaniHamed:2003uy,ArkaniHamed:2005gu}.

\section{An extension: aether SUSY breaking}\label{Ether}
We propose a possible extension to overcome the above difficulties of the kinetic SUSY breaking scenario: It was pointed out that {\it gauged} ghost condensation relaxes the constraint on $M$ to be $M<\mathcal{O}(10^{12}){\rm GeV}$ and also avoids the caustic formation issue~\cite{Cheng:2006us}. In the gauged ghost condensation, we consider gauged shift symmetry $\phi\to \phi-M\xi(x)$, $\xi(x)$ is a U(1) gauge transformation parameter, and introduce an associated U(1) gauge field $A_\mu(x)$, which transforms as $A(x)\to A(x)+\partial_\mu\xi(x)$. As a result, the gauge invariant kinetic term of $\phi$ becomes
\begin{equation}
    \hat{X}=\frac12M^2\left(\partial_\mu\phi/M+A_\mu(x)\right)^2\equiv\frac12 M^2{\cal A}^2.
\end{equation}
In this case, the non-vanishing scalar fluid can be replaced with vector condensation and a simple replacement of $P(\hat{X})$ with $P({\cal A}^2/2)$ reads the ``potential'' for the gauge field ${\cal A}_\mu$. Accordingly, the background solution in non-gauged ghost condensation $X=X_0$ becomes ${\cal A}={\cal A}_0$ in the gauged case. 

As we mentioned above, the problem in non-gauged ghost condensation can be relaxed by the gauging extension. Essentially, in the gauged ghost condensate case, the Goldstone mode $\phi$ more strongly couples to the gauge field rather than gravity, which reduces the modification of gravity. Besides that, we also notice an important feature: Since the scalar $\phi$ can be eaten by the gauge field $A_\mu(x)$, the field excursion of $\phi$ is no longer physical, or in other words, the scalar field is driving in (unobservable) ``gauge'' space. Therefore, such a model can avoid extremely large field excursion problem discussed in the previous section.

Supersymmetric extension of the above idea is straightforwardly done as follows: We introduce a vector multiplet $V$, and the gauge transformations are 
\begin{eqnarray}
    \Phi&\to& \Phi-M\Lambda\\
    V&\to& V+\Lambda+\bar{\Lambda},
\end{eqnarray}
where $\Lambda$ is a gauge transformation parameter superfield. The gauge invariant combination is $\hat{\cal V}=M^{-1}(\Phi+\bar{\Phi}+MV)$. The gauge extension of the Lagrangian~\eqref{PX} is given by
\begin{equation}
    \mathcal{L}=\int d^4\theta \left(M^2K(\hat{\cal V})+\tilde{T}(\hat{\cal A}^2/M^2)D^\alpha\hat{\cal V}D_\alpha\hat{\cal V}\bar{D}_{\dot{\alpha}}\hat{\cal V}\bar{D}^{\dot{\alpha}}\hat{\cal V}\right),\label{PA}
\end{equation}
where we have introduced
\begin{equation}
    \hat{\cal A}_\mu\equiv \frac{1}{4}\bar{\sigma}^\mu_{\alpha\dot\alpha}(D^\alpha\bar{D}^{\dot\alpha}-\bar{D}^{\dot\alpha}D^\alpha)\hat{\cal V}.
\end{equation}
The lowest component of $\hat{\cal A}_\mu$ is ${\cal A}_\mu$, and we note that this is a gauge invariant quantity. These terms provide ``potential'' for ${\cal A}_0$, which causes vector condensation.
The kinetic term of the vector superfield is 
\begin{equation}
    {\cal L}=\frac{1}{4g^2}\int d^2\theta{\cal W}^\alpha{\cal W}_\alpha+{\rm h.c.},
\end{equation}
where ${\cal W}_\alpha=-\frac14 \bar{D}^2D_\alpha \hat{\cal V}$. One can think of this system to be that of a massive vector multiplet $\hat{\cal V}$, and the terms in \eqref{PA} can be regarded as nontrivial potential for massive gauge multiplet, which leads to the vector condensation as mentioned above. We also note that the superpartner scalar $\chi$ can be massive as in the previous case. So far, we have described the system using global SUSY, and one can in principle extend the system to supergravity; however, it would make the analysis more complicated. We leave it for future work. In the following discussion, we discuss couplings to gravity simply by covariantizing the component action.\footnote{Nevertheless, we mention the possible supergravity corrections: One of the concerns of supergravity corrections to bosonic component originates from the correction from auxiliary field in gravity multiplet. We can embed the action~\eqref{PA} into comformal supergravity by making $D_\alpha$ the superconformal spinor derivative~\cite{Kugo:1983mv}. Also, $\hat{V}$ can be identified with a real multiplet with zero conformal weight. Then, the spinor derivative is independent of chiral compensator. Since the auxiliary field in Poincar\'e supergravity multiplet is the F-term of chiral compensator, we expect the supergravity extension does not lead to non-trivial corrections to the bosonic action except minimal couplings to gravity. Indeed, for the ungauged case, there is no such correction, see e.g.~\cite{Aoki:2015eba}. } The vector condensation is achieved as the case of $P(X)$ model. Therefore, we only focus on the dynamics below the condensation scale $M$ in the following.

We briefly describe the low energy dynamics in the gauged ghost condensation model (see~\cite{Cheng:2006us} for details). Below the condensation scale $M$, the 0-th component of the vector ${\cal A}_0$ is integrated out and we are left with the following effective Lagrangian coupled to gravity,
\begin{equation}
    \mathcal{\cal L}=\mathcal{L}_{\rm EH}-\frac{1}{2g^2}(\dot{\cal A}_i-g\epsilon\partial_iH)^2-\frac{1}{4g^2}F_{ij}F^{ij}-\frac12 \alpha(\partial_i{\cal A}^i)^2+\cdots,\label{ggceff}
\end{equation}
where $\epsilon=\frac{M}{\sqrt{2}gM_{\rm pl}}$ and $\alpha$ is a dimensionless constant. Here we have made linearized approximation for gravity around flat spacetime $g_{\mu\nu}=\eta_{\mu\nu}+h_{\mu\nu}$ and $H$ corresponds to canonically normalized gravitational potential $H=M_{\rm pl}h_{00}/\sqrt2 $. We have introduced the last term in~\eqref{ggceff} which would originate e.g. from a higher-order derivative term $-\frac12\alpha(\nabla_\mu{\cal A}^\mu)^2$, which gives the leading order contribution to the longitudinal mode in ${\cal A}_i$. We note that the transverse modes of ${\cal A}_i$ has the standard dispersion relation. Therefore, the non-trivial dispersion relation appears for the longitudinal mode and gravitational potential. Let us parametrize the longitudinal mode as ${\cal A}_i=g\partial_i\sigma$. The gradient term from Einstein-Hilbert action reads ${\mathcal L}_{\rm EH}\supset -\frac12(\partial_iH)^2$ in Newtonian gauge. Thus, we find the quadratic Lagrangian of $\sigma$ and $H$ as
\begin{equation}
    \mathcal{L}=\frac12 (\tilde{\sigma}\ \tilde{H})\left(\begin{array}{cc}
        \omega^2k^2-\alpha g^2k^4 &-{\rm i}\epsilon\omega k^2  \\
        {\rm i}\epsilon\omega k^2 &-k^2(1-\epsilon^2) 
    \end{array}\right)\left(\begin{array}{c}
           \tilde{\sigma}\\
         \tilde{H}
    \end{array}\right).
\end{equation}
Essentially, there is only one propagating mode corresponding to Goldstone mode, and its dispersion relation can be read from the zero of the above kinetic matrix and we find
\begin{equation}
    \omega^2=c^2_sk^2,
\end{equation}
where $c_s^2=\alpha g^2$. Indeed, in the decoupling limit of gravity $M_{\rm pl}\to\infty$ and $\epsilon\to 0$, we are left with a scalar $k\sigma$ with the above dispersion relation. As expected, the longitudinal mode can get the quadratic term $\alpha g^2k^2$ only when $\alpha\neq0$. In the presence of gravity, there is nontrivial mixing between gravitational potential $H$ and $\sigma$, which leads to the two-point function of $H$ of the form,
\begin{equation}
    \langle HH\rangle\sim -\frac{1}{2M_{\rm pl}^2}\frac{1}{k^2}\left(1-\frac{\alpha g^2\epsilon^2k^2}{\omega^2-\alpha g^2(1-\epsilon^2)k^2}\right),
\end{equation}
which leads to the modification of the gravity. Note that in the static limit $\omega\to0$, the above modification becomes just ``renormalization'' of gravitational constant or Planck scale. However, energy source moving relative to the aether background leads to modified profile of gravitational potential, which is constrained by observations discussed below.

The soft SUSY breaking mediation mechanism can be achieved by the replacement $(\Phi+\bar\Phi)\to \hat{\cal V}$. The SUSY breaking can be described by introducing spurion superfield,
\begin{equation}
    v=\langle{\cal V}\rangle=-\theta\sigma^0\bar{\theta}M,
\end{equation}
where $\langle {\cal A}_0\rangle=M$, and we find $v^2=\frac12 M^2\theta^2\bar{\theta}^2$. This is different from the chiral spurion superfield for F-term breaking models $s=\theta^2\langle F\rangle$, but its square is the same, $s\bar{s}=\theta^2\bar{\theta}^2\langle|F|^2\rangle$. We also note that the D-term spurion vector field is $g=\theta^2\bar\theta^2 \langle D\rangle$. Therefore, if the leading order couplings of vector condensate to matter superfields have (accidental) $Z_2$ symmetry $\hat{\cal V}\to-\hat{\cal V}$, the effect of Lorentz violation to the visible sector can be suppressed.  As the kinetic SUSY breaking case, the ``gravity mediation'' can be realized by couplings such as
\begin{equation}
    -\int d^4\theta \frac{M^2}{M_{\rm pl}^2}\hat{\cal V}^2|Q|^2.
\end{equation}
This is the generalization of the coupling~\eqref{softc}, which is reproduced by turning off the gauge superfield $V$. After replacing $\hat{V}$ with the spurion $v$, we find the soft scalar mass to be $m_{Q}^2=M^4/M^2_{\rm pl}$ as the previous case.

Let us discuss the constraints on model parameters. As discussed in~\cite{Cheng:2006us}, the gauged ghost condensation model predicts the modification of the dynamics of gravitational potential, especially for the energy density source moving with velocity $v$ in preferred frame relative to the ghost condensate. As a result, the gravitational potential shows the angular dependence, which can be constrained by the parametrized post-Newtoninan parameter $\alpha_2$, which is related to the model parameters as
\begin{equation}
    \alpha_2=\frac{M^2}{2\alpha g^4(1-\epsilon^2)^2M_{\rm pl}^2},
\end{equation}
The observational constraint $\alpha_2<2\times 10^{-9}$~\cite{Will:2014kxa} leads to the constraint on the scale $M$ as $M<\sqrt{\alpha}g\times10^{14}{\rm GeV}$. The other constraint comes from black hole accretion~\cite{Cheng:2006us}, since the gauged ghost condensation surrounding black hole can make the accretion faster. The observation of XTE J1118+480 leads to $\sqrt{\alpha}M<10^{12}{\rm GeV}$~\cite{Cheng:2006us},
which in our case is translated to the constraint on gravitino mass,
\begin{equation}
    |m_{3/2}|\sim \frac{M^2}{\sqrt{3}M_{\rm pl}}<\mathcal{O}(100){\rm TeV},
\end{equation}
where we have assumed $\alpha\sim g\sim 1$.
If we assume a simple gravity mediation, $m_Q\sim m_{3/2}$, which implies that SUSY partners may be discovered in future collider experiments.\footnote{Recently, a conjecture on the gravitino mass is proposed in \cite{Cribiori:2021gbf}. }

We have to mention necessary fine-tuning of parameters required from the constraints on gravitational wave speed: In the above discussion, we have assumed $-\frac12\alpha(\nabla_\mu{\cal A}^{\mu})^2$, which would be the leading contribution to the dispersion relation of longitudinal mode of ${\cal A}_i$. However, from symmetry viewpoint, $\beta(\nabla_i{\cal A}_j)^2$ can equally contribute, where $\beta$ is a dimensionless coupling constant. However, such a coupling includes $\beta({\langle\cal A}_0\rangle\dot{h}_{ij})^2\sim \beta M^2(\dot{h}_{ij})^2$, which contributes to the dispersion relation of gravitational waves, and changes the sound speed of graviton as quoted in~\cite{Cheng:2006us}. We can make an estimate of the correction to be $|c_s^2-1|^2\sim  \beta M^2/M_{\rm pl}^2$. Since the gravitational wave speed is constrained by the observations of gravitational wave and gamma ray, and the constraint is $|c_s-1|^2\sim 10^{-15}$~\cite{Monitor:2017mdv}. Therefore, we have to require $\sqrt{|\beta|}M/M_{\rm pl}<10^{-15}$. For $M\sim 10^{12}{\rm GeV}$, the constraint on the parameter is $\beta<10^{-9}$, which seems a very unnatural requirement. Unfortunately, we have not found any explanations how to suppress the modification of the graviton dispersion relation. We expect such a requirement would be necessary for various Lorentz violation models unless the violation scale is sufficiently small. In the ``gravity mediation'' discussed above, small Lorentz violation cannot realize sufficiently large mass splitting between SUSY particles and standard model particles.\footnote{Even though stronger couplings between aether field and standard model sector give rise to sufficiently large soft masses with a smaller Lorentz violation. However, of course, direct couplings between standard model sector and aether would be also constrained. Therefore, such a possibility would not be able to reduce the fine-tuning mentioned above.} The other possible constraints would be from Lorentz violation in standard model sector. If we allow various direct couplings between standard model sector and aether sector, more stringent constraints would be expected~\cite{Cheng:2006us}. Regarding Lorentz violation in visible sector, it was pointed out that SUSY suppresses the possible violation~\cite{GrootNibbelink:2004za,Bolokhov:2005cj}. Nevertheless, we expect that in general, couplings to visible sector needs certain amount of fine-tuning. As discussed in~\cite{Cheng:2006us}, if there is a symmetry ${\cal A}_\mu\to -{\cal A}_\mu$ and the aether sector communicates with visible sector via gravity, the violation in visible sector can be rather suppressed, which we assumed in our discussion.

We finally note that the soft mass spectra could be modified if we consider more general cases. For instance, we have assumed that all F-terms and D-terms are vanishing; however, hidden sector coupled to the aether field may lead to non-vanishing F- or D-term of them. In such a case, the soft SUSY mass spectra of standard model sector can differ from the one discussed above. Such a more involved setup might be important to construct realistic models. Note also that, we have assumed ${\cal A}_0\sim M$, which seems rather natural. Also, we have identified the mediation scale $\tilde M$ to be the Planck scale. However, if we construct concrete models (with probably fine-tuned parameters), the scale relation between $M$ and $|m_{3/2}|$ and accordingly the soft mass scale can be different from our estimation. Then, the constraints discussed here may be relaxed.

\section{Conclusion}\label{concl}
We have proposed a new class of SUSY breaking models, in which both Lorentz symmetry and SUSY are spontaneously broken. As we have shown, scalar fluid, or vector condensation may play the role of F-term, and soft SUSY breaking in visible sector can be realized in a way similar to F-term breaking scenarios. However, as we have seen in the kinetic SUSY breaking scenario, sufficiently large velocity is required for almost whole history of the Universe, which leads to extremely large field excursion. We have also proposed the improved model, aether SUSY breaking, where the scalar is eaten by a gauge field. Then the scalar drives in ``gauge space'' and the notion of field excursion disappears, which solves the large field excursion problem or does not require exact global symmetry. In both cases, the Lorentz symmetry needs to be broken at a very high scale.

Our model discussed in this paper is one of the possibilities of more general class of Lorentz violating SUSY breaking scenarios. The most important point of our proposal is that SUSY breaking patterns for realistic model buildings is not restricted only within F-term or D-term breaking scenarios. Investigation of our new SUSY breaking scenario from UV theories such as string theory would be interesting. We note that realization of (gauged) ghost condensation is discussed e.g. in~\cite{Mukohyama:2006mm}. The recently proposed ``revolving D-brane'' model~\cite{Iso:2015mva} may have similarity to our model since the motion of D-brane leads to Lorentz violation and accordingly SUSY is spontaneously broken.

One of the interesting features of our model is that the Lorentz violation scale is tightly related to SUSY breaking scale or soft SUSY mass scale. Therefore, the model is testable either by collider experiments or (astrophysical) modified gravity tests. Specifically, the constraints discussed within our simple model implies the SUSY breaking scale to be less than ${\cal O}(100)$ TeV. Therefore, future collider experiments can test our model. 

There are still various issues we should address. In this work, we have shown a specific SUSY breaking setup; however, we have not specified possible matter couplings induced by Lorentz violation. The Lorentz violation in the standard model sector was studied in various literature (see e.g.~\cite{Colladay:1996iz,Coleman:1997xq,Coleman:1998ti,Colladay:1998fq} for earlier works.) In order to investigate such an issue, we need to specify the coupling between matter sector and aether superfield. As discussed in~\cite{Cheng:2006us}, discrete symmetry may suppress the leading order Lorentz violation effects on the standard model, and SUSY may further suppress the violation effects~\cite{GrootNibbelink:2004za,Bolokhov:2005cj}. It would be worth studying more details of phenomenological side of this scenario. It would also be important to discuss the connection to very early Universe such as inflationary stage. We leave these study for future work.

\section*{acknowledgement}
I would like to thank Jun'ichi Yokoyama for useful discussions and comments. This work is supported by JSPS KAKENHI, Grant-in-Aid for JSPS Fellows JP19J00494.

\bibliographystyle{JHEP}
\bibliography{ref}

\providecommand{\href}[2]{#2}\begingroup\raggedright\begin{thebibliography}{10}

\bibitem{Witten:1982df}
E.~Witten, \emph{{Constraints on Supersymmetry Breaking}},
  \href{https://doi.org/10.1016/0550-3213(82)90071-2}{\emph{Nucl. Phys. B}
  {\bfseries 202} (1982) 253}.

\bibitem{Affleck:1983mk}
I.~Affleck, M.~Dine and N.~Seiberg, \emph{{Dynamical Supersymmetry Breaking in
  Supersymmetric QCD}},
  \href{https://doi.org/10.1016/0550-3213(84)90058-0}{\emph{Nucl. Phys. B}
  {\bfseries 241} (1984) 493}.

\bibitem{Affleck:1984xz}
I.~Affleck, M.~Dine and N.~Seiberg, \emph{{Dynamical Supersymmetry Breaking in
  Four-Dimensions and Its Phenomenological Implications}},
  \href{https://doi.org/10.1016/0550-3213(85)90408-0}{\emph{Nucl. Phys. B}
  {\bfseries 256} (1985) 557}.

\bibitem{Affleck:1984uz}
I.~Affleck, M.~Dine and N.~Seiberg, \emph{{Calculable Nonperturbative
  Supersymmetry Breaking}},
  \href{https://doi.org/10.1103/PhysRevLett.52.1677}{\emph{Phys. Rev. Lett.}
  {\bfseries 52} (1984) 1677}.

\bibitem{Izawa:1996pk}
K.-I. Izawa and T.~Yanagida, \emph{{Dynamical supersymmetry breaking in vector
  - like gauge theories}},
  \href{https://doi.org/10.1143/PTP.95.829}{\emph{Prog. Theor. Phys.}
  {\bfseries 95} (1996) 829}
  [\href{https://arxiv.org/abs/hep-th/9602180}{{\ttfamily hep-th/9602180}}].

\bibitem{Intriligator:1996pu}
K.~A. Intriligator and S.~D. Thomas, \emph{{Dynamical supersymmetry breaking on
  quantum moduli spaces}},
  \href{https://doi.org/10.1016/0550-3213(96)00261-1}{\emph{Nucl. Phys. B}
  {\bfseries 473} (1996) 121}
  [\href{https://arxiv.org/abs/hep-th/9603158}{{\ttfamily hep-th/9603158}}].

\bibitem{Intriligator:2006dd}
K.~A. Intriligator, N.~Seiberg and D.~Shih, \emph{{Dynamical SUSY breaking in
  meta-stable vacua}},
  \href{https://doi.org/10.1088/1126-6708/2006/04/021}{\emph{JHEP} {\bfseries
  04} (2006) 021} [\href{https://arxiv.org/abs/hep-th/0602239}{{\ttfamily
  hep-th/0602239}}].

\bibitem{Kachru:2003aw}
S.~Kachru, R.~Kallosh, A.~D. Linde and S.~P. Trivedi, \emph{{De Sitter vacua in
  string theory}},
  \href{https://doi.org/10.1103/PhysRevD.68.046005}{\emph{Phys. Rev. D}
  {\bfseries 68} (2003) 046005}
  [\href{https://arxiv.org/abs/hep-th/0301240}{{\ttfamily hep-th/0301240}}].

\bibitem{Balasubramanian:2005zx}
V.~Balasubramanian, P.~Berglund, J.~P. Conlon and F.~Quevedo,
  \emph{{Systematics of moduli stabilisation in Calabi-Yau flux
  compactifications}},
  \href{https://doi.org/10.1088/1126-6708/2005/03/007}{\emph{JHEP} {\bfseries
  03} (2005) 007} [\href{https://arxiv.org/abs/hep-th/0502058}{{\ttfamily
  hep-th/0502058}}].

\bibitem{Conlon:2005ki}
J.~P. Conlon, F.~Quevedo and K.~Suruliz, \emph{{Large-volume flux
  compactifications: Moduli spectrum and D3/D7 soft supersymmetry breaking}},
  \href{https://doi.org/10.1088/1126-6708/2005/08/007}{\emph{JHEP} {\bfseries
  08} (2005) 007} [\href{https://arxiv.org/abs/hep-th/0505076}{{\ttfamily
  hep-th/0505076}}].

\bibitem{Kachru:2002gs}
S.~Kachru, J.~Pearson and H.~L. Verlinde, \emph{{Brane / flux annihilation and
  the string dual of a nonsupersymmetric field theory}},
  \href{https://doi.org/10.1088/1126-6708/2002/06/021}{\emph{JHEP} {\bfseries
  06} (2002) 021} [\href{https://arxiv.org/abs/hep-th/0112197}{{\ttfamily
  hep-th/0112197}}].

\bibitem{McGuirk:2012sb}
P.~McGuirk, G.~Shiu and F.~Ye, \emph{{Soft branes in supersymmetry-breaking
  backgrounds}}, \href{https://doi.org/10.1007/JHEP07(2012)188}{\emph{JHEP}
  {\bfseries 07} (2012) 188} [\href{https://arxiv.org/abs/1206.0754}{{\ttfamily
  1206.0754}}].

\bibitem{Kallosh:2014wsa}
R.~Kallosh and T.~Wrase, \emph{{Emergence of Spontaneously Broken Supersymmetry
  on an Anti-D3-Brane in KKLT dS Vacua}},
  \href{https://doi.org/10.1007/JHEP12(2014)117}{\emph{JHEP} {\bfseries 12}
  (2014) 117} [\href{https://arxiv.org/abs/1411.1121}{{\ttfamily 1411.1121}}].

\bibitem{Bergshoeff:2015jxa}
E.~A. Bergshoeff, K.~Dasgupta, R.~Kallosh, A.~Van~Proeyen and T.~Wrase,
  \emph{{$ \overline{\mathrm{D}3} $ and dS}},
  \href{https://doi.org/10.1007/JHEP05(2015)058}{\emph{JHEP} {\bfseries 05}
  (2015) 058} [\href{https://arxiv.org/abs/1502.07627}{{\ttfamily
  1502.07627}}].

\bibitem{Aparicio:2015psl}
L.~Aparicio, F.~Quevedo and R.~Valandro, \emph{{Moduli Stabilisation with
  Nilpotent Goldstino: Vacuum Structure and SUSY Breaking}},
  \href{https://doi.org/10.1007/JHEP03(2016)036}{\emph{JHEP} {\bfseries 03}
  (2016) 036} [\href{https://arxiv.org/abs/1511.08105}{{\ttfamily
  1511.08105}}].

\bibitem{Vercnocke:2016fbt}
B.~Vercnocke and T.~Wrase, \emph{{Constrained superfields from an anti-D3-brane
  in KKLT}}, \href{https://doi.org/10.1007/JHEP08(2016)132}{\emph{JHEP}
  {\bfseries 08} (2016) 132}
  [\href{https://arxiv.org/abs/1605.03961}{{\ttfamily 1605.03961}}].

\bibitem{Kallosh:2016aep}
R.~Kallosh, B.~Vercnocke and T.~Wrase, \emph{{String Theory Origin of
  Constrained Multiplets}},
  \href{https://doi.org/10.1007/JHEP09(2016)063}{\emph{JHEP} {\bfseries 09}
  (2016) 063} [\href{https://arxiv.org/abs/1606.09245}{{\ttfamily
  1606.09245}}].

\bibitem{GarciadelMoral:2017vnz}
M.~P. Garcia~del Moral, S.~Parameswaran, N.~Quiroz and I.~Zavala,
  \emph{{Anti-D3 branes and moduli in non-linear supergravity}},
  \href{https://doi.org/10.1007/JHEP10(2017)185}{\emph{JHEP} {\bfseries 10}
  (2017) 185} [\href{https://arxiv.org/abs/1707.07059}{{\ttfamily
  1707.07059}}].

\bibitem{Aalsma:2018pll}
L.~Aalsma, M.~Tournoy, J.~P. Van Der~Schaar and B.~Vercnocke,
  \emph{{Supersymmetric embedding of antibrane polarization}},
  \href{https://doi.org/10.1103/PhysRevD.98.086019}{\emph{Phys. Rev. D}
  {\bfseries 98} (2018) 086019}
  [\href{https://arxiv.org/abs/1807.03303}{{\ttfamily 1807.03303}}].

\bibitem{Cribiori:2019hod}
N.~Cribiori, C.~Roupec, T.~Wrase and Y.~Yamada, \emph{{Supersymmetric
  anti-D3-brane action in the Kachru-Kallosh-Linde-Trivedi setup}},
  \href{https://doi.org/10.1103/PhysRevD.100.066001}{\emph{Phys. Rev. D}
  {\bfseries 100} (2019) 066001}
  [\href{https://arxiv.org/abs/1906.07727}{{\ttfamily 1906.07727}}].

\bibitem{Volkov:1972jx}
D.~V. Volkov and V.~P. Akulov, \emph{{Possible universal neutrino
  interaction}}, {\emph{JETP Lett.} {\bfseries 16} (1972) 438}.

\bibitem{Volkov:1973ix}
D.~V. Volkov and V.~P. Akulov, \emph{{Is the Neutrino a Goldstone Particle?}},
  \href{https://doi.org/10.1016/0370-2693(73)90490-5}{\emph{Phys. Lett. B}
  {\bfseries 46} (1973) 109}.

\bibitem{Rocek:1978nb}
M.~Rocek, \emph{{Linearizing the Volkov-Akulov Model}},
  \href{https://doi.org/10.1103/PhysRevLett.41.451}{\emph{Phys. Rev. Lett.}
  {\bfseries 41} (1978) 451}.

\bibitem{Ivanov:1978mx}
E.~A. Ivanov and A.~A. Kapustnikov, \emph{{General Relationship Between Linear
  and Nonlinear Realizations of Supersymmetry}},
  \href{https://doi.org/10.1088/0305-4470/11/12/005}{\emph{J. Phys. A}
  {\bfseries 11} (1978) 2375}.

\bibitem{Lindstrom:1979kq}
U.~Lindstrom and M.~Rocek, \emph{{CONSTRAINED LOCAL SUPERFIELDS}},
  \href{https://doi.org/10.1103/PhysRevD.19.2300}{\emph{Phys. Rev. D}
  {\bfseries 19} (1979) 2300}.

\bibitem{Casalbuoni:1988xh}
R.~Casalbuoni, S.~De~Curtis, D.~Dominici, F.~Feruglio and R.~Gatto,
  \emph{{Nonlinear Realization of Supersymmetry Algebra From Supersymmetric
  Constraint}}, \href{https://doi.org/10.1016/0370-2693(89)90788-0}{\emph{Phys.
  Lett. B} {\bfseries 220} (1989) 569}.

\bibitem{Komargodski:2009rz}
Z.~Komargodski and N.~Seiberg, \emph{{From Linear SUSY to Constrained
  Superfields}},
  \href{https://doi.org/10.1088/1126-6708/2009/09/066}{\emph{JHEP} {\bfseries
  09} (2009) 066} [\href{https://arxiv.org/abs/0907.2441}{{\ttfamily
  0907.2441}}].

\bibitem{Jacobson:2000xp}
T.~Jacobson and D.~Mattingly, \emph{{Gravity with a dynamical preferred
  frame}}, \href{https://doi.org/10.1103/PhysRevD.64.024028}{\emph{Phys. Rev.
  D} {\bfseries 64} (2001) 024028}
  [\href{https://arxiv.org/abs/gr-qc/0007031}{{\ttfamily gr-qc/0007031}}].

\bibitem{Jacobson:2008aj}
T.~Jacobson, \emph{{Einstein-aether gravity: A Status report}},
  \href{https://doi.org/10.22323/1.043.0020}{\emph{PoS} {\bfseries QG-PH}
  (2007) 020} [\href{https://arxiv.org/abs/0801.1547}{{\ttfamily 0801.1547}}].

\bibitem{Palti:2019pca}
E.~Palti, \emph{{The Swampland: Introduction and Review}},
  \href{https://doi.org/10.1002/prop.201900037}{\emph{Fortsch. Phys.}
  {\bfseries 67} (2019) 1900037}
  [\href{https://arxiv.org/abs/1903.06239}{{\ttfamily 1903.06239}}].

\bibitem{Obied:2018sgi}
G.~Obied, H.~Ooguri, L.~Spodyneiko and C.~Vafa, \emph{{De Sitter Space and the
  Swampland}},  \href{https://arxiv.org/abs/1806.08362}{{\ttfamily
  1806.08362}}.

\bibitem{Garg:2018reu}
S.~K. Garg and C.~Krishnan, \emph{{Bounds on Slow Roll and the de Sitter
  Swampland}}, \href{https://doi.org/10.1007/JHEP11(2019)075}{\emph{JHEP}
  {\bfseries 11} (2019) 075}
  [\href{https://arxiv.org/abs/1807.05193}{{\ttfamily 1807.05193}}].

\bibitem{Ooguri:2018wrx}
H.~Ooguri, E.~Palti, G.~Shiu and C.~Vafa, \emph{{Distance and de Sitter
  Conjectures on the Swampland}},
  \href{https://doi.org/10.1016/j.physletb.2018.11.018}{\emph{Phys. Lett. B}
  {\bfseries 788} (2019) 180}
  [\href{https://arxiv.org/abs/1810.05506}{{\ttfamily 1810.05506}}].

\bibitem{Wess:1992cp}
J.~Wess and J.~Bagger, \emph{{Supersymmetry and supergravity}}. Princeton
  University Press, Princeton, NJ, USA, 1992.

\bibitem{Cribiori:2017laj}
N.~Cribiori, F.~Farakos, M.~Tournoy and A.~van Proeyen, \emph{{Fayet-Iliopoulos
  terms in supergravity without gauged R-symmetry}},
  \href{https://doi.org/10.1007/JHEP04(2018)032}{\emph{JHEP} {\bfseries 04}
  (2018) 032} [\href{https://arxiv.org/abs/1712.08601}{{\ttfamily
  1712.08601}}].

\bibitem{Antoniadis:2018cpq}
I.~Antoniadis, A.~Chatrabhuti, H.~Isono and R.~Knoops,
  \emph{{Fayet\textendash{}Iliopoulos terms in supergravity and D-term
  inflation}}, \href{https://doi.org/10.1140/epjc/s10052-018-5861-6}{\emph{Eur.
  Phys. J. C} {\bfseries 78} (2018) 366}
  [\href{https://arxiv.org/abs/1803.03817}{{\ttfamily 1803.03817}}].

\bibitem{Kuzenko:2018jlz}
S.~M. Kuzenko, \emph{{Taking a vector supermultiplet apart: Alternative
  Fayet\textendash{}Iliopoulos-type terms}},
  \href{https://doi.org/10.1016/j.physletb.2018.04.051}{\emph{Phys. Lett. B}
  {\bfseries 781} (2018) 723}
  [\href{https://arxiv.org/abs/1801.04794}{{\ttfamily 1801.04794}}].

\bibitem{Martin:1997ns}
S.~P. Martin, \emph{{A Supersymmetry primer}},
  \href{https://doi.org/10.1142/9789812839657_0001}{\emph{Adv. Ser. Direct.
  High Energy Phys.} {\bfseries 21} (2010) 1}
  [\href{https://arxiv.org/abs/hep-ph/9709356}{{\ttfamily hep-ph/9709356}}].

\bibitem{Randall:1998uk}
L.~Randall and R.~Sundrum, \emph{{Out of this world supersymmetry breaking}},
  \href{https://doi.org/10.1016/S0550-3213(99)00359-4}{\emph{Nucl. Phys. B}
  {\bfseries 557} (1999) 79}
  [\href{https://arxiv.org/abs/hep-th/9810155}{{\ttfamily hep-th/9810155}}].

\bibitem{Giudice:1998xp}
G.~F. Giudice, M.~A. Luty, H.~Murayama and R.~Rattazzi, \emph{{Gaugino mass
  without singlets}},
  \href{https://doi.org/10.1088/1126-6708/1998/12/027}{\emph{JHEP} {\bfseries
  12} (1998) 027} [\href{https://arxiv.org/abs/hep-ph/9810442}{{\ttfamily
  hep-ph/9810442}}].

\bibitem{Bagger:1999rd}
J.~A. Bagger, T.~Moroi and E.~Poppitz, \emph{{Anomaly mediation in supergravity
  theories}}, \href{https://doi.org/10.1088/1126-6708/2000/04/009}{\emph{JHEP}
  {\bfseries 04} (2000) 009}
  [\href{https://arxiv.org/abs/hep-th/9911029}{{\ttfamily hep-th/9911029}}].

\bibitem{Koehn:2012te}
M.~Koehn, J.-L. Lehners and B.~Ovrut, \emph{{Ghost condensate in $N=1$
  supergravity}}, \href{https://doi.org/10.1103/PhysRevD.87.065022}{\emph{Phys.
  Rev. D} {\bfseries 87} (2013) 065022}
  [\href{https://arxiv.org/abs/1212.2185}{{\ttfamily 1212.2185}}].

\bibitem{Nitta:2017yuf}
M.~Nitta, S.~Sasaki and R.~Yokokura, \emph{{Supersymmetry Breaking in Spatially
  Modulated Vacua}},
  \href{https://doi.org/10.1103/PhysRevD.96.105022}{\emph{Phys. Rev. D}
  {\bfseries 96} (2017) 105022}
  [\href{https://arxiv.org/abs/1706.05232}{{\ttfamily 1706.05232}}].

\bibitem{Gudnason:2018nof}
S.~Bjarke~Gudnason, M.~Nitta, S.~Sasaki and R.~Yokokura, \emph{{Supersymmetry
  breaking and ghost Goldstino in modulated vacua}},
  \href{https://doi.org/10.1103/PhysRevD.99.045012}{\emph{Phys. Rev. D}
  {\bfseries 99} (2019) 045012}
  [\href{https://arxiv.org/abs/1812.09078}{{\ttfamily 1812.09078}}].

\bibitem{Katz:2006rx}
A.~Katz and Y.~Shadmi, \emph{{Lorentz Violation and Superpartner Masses}},
  \href{https://doi.org/10.1103/PhysRevD.74.115021}{\emph{Phys. Rev. D}
  {\bfseries 74} (2006) 115021}
  [\href{https://arxiv.org/abs/hep-ph/0605210}{{\ttfamily hep-ph/0605210}}].

\bibitem{Chiba:1999ka}
T.~Chiba, T.~Okabe and M.~Yamaguchi, \emph{{Kinetically driven quintessence}},
  \href{https://doi.org/10.1103/PhysRevD.62.023511}{\emph{Phys. Rev. D}
  {\bfseries 62} (2000) 023511}
  [\href{https://arxiv.org/abs/astro-ph/9912463}{{\ttfamily
  astro-ph/9912463}}].

\bibitem{ArkaniHamed:2003uy}
N.~Arkani-Hamed, H.-C. Cheng, M.~A. Luty and S.~Mukohyama, \emph{{Ghost
  condensation and a consistent infrared modification of gravity}},
  \href{https://doi.org/10.1088/1126-6708/2004/05/074}{\emph{JHEP} {\bfseries
  05} (2004) 074} [\href{https://arxiv.org/abs/hep-th/0312099}{{\ttfamily
  hep-th/0312099}}].

\bibitem{Khoury:2010gb}
J.~Khoury, J.-L. Lehners and B.~Ovrut, \emph{{Supersymmetric P(X,$\phi$) and
  the Ghost Condensate}},
  \href{https://doi.org/10.1103/PhysRevD.83.125031}{\emph{Phys. Rev. D}
  {\bfseries 83} (2011) 125031}
  [\href{https://arxiv.org/abs/1012.3748}{{\ttfamily 1012.3748}}].

\bibitem{Koehn:2012ar}
M.~Koehn, J.-L. Lehners and B.~A. Ovrut, \emph{{Higher-Derivative Chiral
  Superfield Actions Coupled to N=1 Supergravity}},
  \href{https://doi.org/10.1103/PhysRevD.86.085019}{\emph{Phys. Rev. D}
  {\bfseries 86} (2012) 085019}
  [\href{https://arxiv.org/abs/1207.3798}{{\ttfamily 1207.3798}}].

\bibitem{Aoki:2014pna}
S.~Aoki and Y.~Yamada, \emph{{Inflation in supergravity without K\"ahler
  potential}}, \href{https://doi.org/10.1103/PhysRevD.90.127701}{\emph{Phys.
  Rev. D} {\bfseries 90} (2014) 127701}
  [\href{https://arxiv.org/abs/1409.4183}{{\ttfamily 1409.4183}}].

\bibitem{Aoki:2015eba}
S.~Aoki and Y.~Yamada, \emph{{Impacts of supersymmetric higher derivative terms
  on inflation models in supergravity}},
  \href{https://doi.org/10.1088/1475-7516/2015/07/020}{\emph{JCAP} {\bfseries
  07} (2015) 020} [\href{https://arxiv.org/abs/1504.07023}{{\ttfamily
  1504.07023}}].

\bibitem{Farakos:2021mwt}
F.~Farakos, \emph{{On the F-term problem and quintessence supersymmetry
  breaking}},  \href{https://arxiv.org/abs/2101.05759}{{\ttfamily 2101.05759}}.

\bibitem{DEramo:2012vvz}
F.~D'Eramo, J.~Thaler and Z.~Thomas, \emph{{The Two Faces of Anomaly
  Mediation}}, \href{https://doi.org/10.1007/JHEP06(2012)151}{\emph{JHEP}
  {\bfseries 06} (2012) 151} [\href{https://arxiv.org/abs/1202.1280}{{\ttfamily
  1202.1280}}].

\bibitem{DEramo:2013dzi}
F.~D'Eramo, J.~Thaler and Z.~Thomas, \emph{{Anomaly Mediation from Unbroken
  Supergravity}}, \href{https://doi.org/10.1007/JHEP09(2013)125}{\emph{JHEP}
  {\bfseries 09} (2013) 125} [\href{https://arxiv.org/abs/1307.3251}{{\ttfamily
  1307.3251}}].

\bibitem{Ooguri:2006in}
H.~Ooguri and C.~Vafa, \emph{{On the Geometry of the String Landscape and the
  Swampland}},
  \href{https://doi.org/10.1016/j.nuclphysb.2006.10.033}{\emph{Nucl. Phys. B}
  {\bfseries 766} (2007) 21}
  [\href{https://arxiv.org/abs/hep-th/0605264}{{\ttfamily hep-th/0605264}}].

\bibitem{Kallosh:1995hi}
R.~Kallosh, A.~D. Linde, D.~A. Linde and L.~Susskind, \emph{{Gravity and global
  symmetries}}, \href{https://doi.org/10.1103/PhysRevD.52.912}{\emph{Phys. Rev.
  D} {\bfseries 52} (1995) 912}
  [\href{https://arxiv.org/abs/hep-th/9502069}{{\ttfamily hep-th/9502069}}].

\bibitem{Banks:2010zn}
T.~Banks and N.~Seiberg, \emph{{Symmetries and Strings in Field Theory and
  Gravity}}, \href{https://doi.org/10.1103/PhysRevD.83.084019}{\emph{Phys. Rev.
  D} {\bfseries 83} (2011) 084019}
  [\href{https://arxiv.org/abs/1011.5120}{{\ttfamily 1011.5120}}].

\bibitem{Harlow:2018jwu}
D.~Harlow and H.~Ooguri, \emph{{Constraints on Symmetries from Holography}},
  \href{https://doi.org/10.1103/PhysRevLett.122.191601}{\emph{Phys. Rev. Lett.}
  {\bfseries 122} (2019) 191601}
  [\href{https://arxiv.org/abs/1810.05337}{{\ttfamily 1810.05337}}].

\bibitem{Harlow:2018tng}
D.~Harlow and H.~Ooguri, \emph{{Symmetries in quantum field theory and quantum
  gravity}},  \href{https://arxiv.org/abs/1810.05338}{{\ttfamily 1810.05338}}.

\bibitem{ArkaniHamed:2005gu}
N.~Arkani-Hamed, H.-C. Cheng, M.~A. Luty, S.~Mukohyama and T.~Wiseman,
  \emph{{Dynamics of gravity in a Higgs phase}},
  \href{https://doi.org/10.1088/1126-6708/2007/01/036}{\emph{JHEP} {\bfseries
  01} (2007) 036} [\href{https://arxiv.org/abs/hep-ph/0507120}{{\ttfamily
  hep-ph/0507120}}].

\bibitem{Cheng:2006us}
H.-C. Cheng, M.~A. Luty, S.~Mukohyama and J.~Thaler, \emph{{Spontaneous Lorentz
  breaking at high energies}},
  \href{https://doi.org/10.1088/1126-6708/2006/05/076}{\emph{JHEP} {\bfseries
  05} (2006) 076} [\href{https://arxiv.org/abs/hep-th/0603010}{{\ttfamily
  hep-th/0603010}}].

\bibitem{Kugo:1983mv}
T.~Kugo and S.~Uehara, \emph{{$N=1$ Superconformal Tensor Calculus: Multiplets
  With External Lorentz Indices and Spinor Derivative Operators}},
  \href{https://doi.org/10.1143/PTP.73.235}{\emph{Prog. Theor. Phys.}
  {\bfseries 73} (1985) 235}.

\bibitem{Will:2014kxa}
C.~M. Will, \emph{{The Confrontation between General Relativity and
  Experiment}}, \href{https://doi.org/10.12942/lrr-2014-4}{\emph{Living Rev.
  Rel.} {\bfseries 17} (2014) 4}
  [\href{https://arxiv.org/abs/1403.7377}{{\ttfamily 1403.7377}}].

\bibitem{Cribiori:2021gbf}
N.~Cribiori, D.~L\"ust and M.~Scalisi, \emph{{The gravitino and the
  swampland}}, \href{https://doi.org/10.1007/JHEP06(2021)071}{\emph{JHEP}
  {\bfseries 06} (2021) 071}
  [\href{https://arxiv.org/abs/2104.08288}{{\ttfamily 2104.08288}}].

\bibitem{Monitor:2017mdv}
{\scshape LIGO Scientific, Virgo, Fermi-GBM, INTEGRAL} collaboration,
  \emph{{Gravitational Waves and Gamma-rays from a Binary Neutron Star Merger:
  GW170817 and GRB 170817A}},
  \href{https://doi.org/10.3847/2041-8213/aa920c}{\emph{Astrophys. J. Lett.}
  {\bfseries 848} (2017) L13}
  [\href{https://arxiv.org/abs/1710.05834}{{\ttfamily 1710.05834}}].

\bibitem{GrootNibbelink:2004za}
S.~Groot~Nibbelink and M.~Pospelov, \emph{{Lorentz violation in supersymmetric
  field theories}},
  \href{https://doi.org/10.1103/PhysRevLett.94.081601}{\emph{Phys. Rev. Lett.}
  {\bfseries 94} (2005) 081601}
  [\href{https://arxiv.org/abs/hep-ph/0404271}{{\ttfamily hep-ph/0404271}}].

\bibitem{Bolokhov:2005cj}
P.~A. Bolokhov, S.~Groot~Nibbelink and M.~Pospelov, \emph{{Lorentz violating
  supersymmetric quantum electrodynamics}},
  \href{https://doi.org/10.1103/PhysRevD.72.015013}{\emph{Phys. Rev. D}
  {\bfseries 72} (2005) 015013}
  [\href{https://arxiv.org/abs/hep-ph/0505029}{{\ttfamily hep-ph/0505029}}].

\bibitem{Mukohyama:2006mm}
S.~Mukohyama, \emph{{Towards a Higgs phase of gravity in string theory}},
  \href{https://doi.org/10.1088/1126-6708/2007/05/048}{\emph{JHEP} {\bfseries
  05} (2007) 048} [\href{https://arxiv.org/abs/hep-th/0610254}{{\ttfamily
  hep-th/0610254}}].

\bibitem{Iso:2015mva}
S.~Iso and N.~Kitazawa, \emph{{Revolving D-branes and Spontaneous Gauge
  Symmetry Breaking}}, \href{https://doi.org/10.1093/ptep/ptv157}{\emph{PTEP}
  {\bfseries 2015} (2015) 123B01}
  [\href{https://arxiv.org/abs/1507.04834}{{\ttfamily 1507.04834}}].

\bibitem{Colladay:1996iz}
D.~Colladay and V.~A. Kostelecky, \emph{{CPT violation and the standard
  model}}, \href{https://doi.org/10.1103/PhysRevD.55.6760}{\emph{Phys. Rev. D}
  {\bfseries 55} (1997) 6760}
  [\href{https://arxiv.org/abs/hep-ph/9703464}{{\ttfamily hep-ph/9703464}}].

\bibitem{Coleman:1997xq}
S.~R. Coleman and S.~L. Glashow, \emph{{Cosmic ray and neutrino tests of
  special relativity}},
  \href{https://doi.org/10.1016/S0370-2693(97)00638-2}{\emph{Phys. Lett. B}
  {\bfseries 405} (1997) 249}
  [\href{https://arxiv.org/abs/hep-ph/9703240}{{\ttfamily hep-ph/9703240}}].

\bibitem{Coleman:1998ti}
S.~R. Coleman and S.~L. Glashow, \emph{{High-energy tests of Lorentz
  invariance}}, \href{https://doi.org/10.1103/PhysRevD.59.116008}{\emph{Phys.
  Rev. D} {\bfseries 59} (1999) 116008}
  [\href{https://arxiv.org/abs/hep-ph/9812418}{{\ttfamily hep-ph/9812418}}].

\bibitem{Colladay:1998fq}
D.~Colladay and V.~A. Kostelecky, \emph{{Lorentz violating extension of the
  standard model}},
  \href{https://doi.org/10.1103/PhysRevD.58.116002}{\emph{Phys. Rev. D}
  {\bfseries 58} (1998) 116002}
  [\href{https://arxiv.org/abs/hep-ph/9809521}{{\ttfamily hep-ph/9809521}}].

\end{thebibliography}\endgroup
\end{document}